\documentclass[10pt,twocolumn,superscriptaddress,aps,prb,balancelastpage,longbibliography]{revtex4-1}
\usepackage{lmodern}
\usepackage{graphicx}
\usepackage{braket}
\usepackage{amsfonts, amsmath, amsthm, amssymb}
\usepackage{bbold}
\usepackage{subfigure}
\usepackage{hyperref}
\hypersetup{colorlinks=true,linkcolor=blue,citecolor=blue,urlcolor=blue}
\def\equationautorefname~#1\null{(#1)\null}
\usepackage{cleveref}
\usepackage[utf8]{inputenc}
\usepackage[T1]{fontenc}
\Crefname{figure}{Fig.}{Figs.}

\makeatletter
\newsavebox{\@brx}
\newcommand{\llangle}[1][]{\savebox{\@brx}{\(\m@th{#1\langle}\)}%
  \mathopen{\copy\@brx\kern-0.5\wd\@brx\usebox{\@brx}}}
\newcommand{\rrangle}[1][]{\savebox{\@brx}{\(\m@th{#1\rangle}\)}%
  \mathclose{\copy\@brx\kern-0.5\wd\@brx\usebox{\@brx}}}
\makeatother
\begin{document}
\title{Structural and Dynamical Fingerprints of the Anomalous Dielectric Properties of Water Under Confinement}
\author{Iman Ahmadabadi} 
\altaffiliation{Correspondence to \href{mailto:imanahmadabadi@physics.sharif.edu}{imanahmadabadi@physics.sharif.edu}}
\affiliation{Department of Physics, Sharif University of Technology, Tehran 14588-89694, Iran}

\author{Ali Esfandiar}
\affiliation{Department of Physics, Sharif University of Technology, Tehran 14588-89694, Iran}
\author{Ali Hassanali} 
\affiliation{The Abdus Salam International Centre for Theoretical Physics, Strada Costiera 11, 34151 Trieste, Italy}

\author{Mohammad Reza Ejtehadi} 
\altaffiliation{Correspondence to \href{mailto:ejtehadi@sharif.edu}{ejtehadi@sharif.edu}}
\affiliation{Department of Physics,
	Sharif University of Technology, Tehran 14588-89694, Iran}
	\homepage[]{Your web page}
	\thanks{}
	\altaffiliation{}

\begin{abstract}
There is a long-standing question about the molecular configuration of interfacial water molecules in the proximity of solid surfaces, particularly carbon atoms which plays a crucial role in electrochemistry and biology. In this study, the dielectric, structural and dynamical properties of confined water placed between two parallel graphene walls at different inter distances from the Angstrom scale to few tens of nanometer have been investigated using molecular dynamics. For dielectric properties of water, we show that the perpendicular component of water dielectric constant drastically decreases under sub $2\;nm$ spatial confinement. The achieved dielectric constant data through linear response and fluctuation-dissipation theory, are consistent with recent reported experimental results.\cite{fumagalli2018anomalously} By determining the charge density as well as fluctuations in the number of atoms, we provide a molecular rationale for the behavior of perpendicular dielectric response function. We also interpret the behavior of the dielectric response in terms of the presence of dangling O-H bonds of waters. By examining the residence time and lateral diffusion constant of water under confinement, we reveal that the water molecules tend to keep their hydrogen bond networks at the interface of water-graphene. We also found consistency between lateral diffusion and z-component of variance in the center of mass of the system as a function of confinement.
\end{abstract}
	
\pacs{to be completed}

\maketitle
\nocite{*}
	
\section{Introduction}	
	
It is well known that the behavior of interfacial water governs both its rheology and physical properties in a wide range of phenomena in nature.\cite{chandler2005interfaces,ball2008water,franks2007water} To understand the interaction and structure of water in contact with solid surfaces, various experimental and theoretical efforts have been applied in the literature.\cite{hummer2001water,algara2015square,chaban2012confinement,secchi2016massive,sugahara2019negative,breynaert2020water,webber2010studies,korb1999anomalous,mallamace2012dynamical,zanotti1999relaxational,jansson2003dynamics,gilijamse2005dynamics,xu2011there,harrach2014structure,faraone2004fragile,rasaiah2008water,senapati2001dielectric,chaban2012confinement} One of the most studied systems in this regard, is the behavior of water near hydrophobic interfaces specifically, near graphene.\cite{breynaert2020water,hummer2001water,algara2015square,chaban2012confinement,secchi2016massive,sugahara2019negative,webber2010studies,algara2015square,sugahara2019negative,senapati2001dielectric,rasaiah2008water} Due to the obvious general relevance, the graphene/water interface has been an attractive system to simulate and study the dynamics of water.\cite{breynaert2020water,hummer2001water,algara2015square,chaban2012confinement,secchi2016massive,algara2015square,sugahara2019negative,chaban2012confinement} This has numerous practical implications, in particular, the presence of water in ultra-narrow slits and membranes in biology and porous electrodes in electrochemistry, make the graphene-water system the subject of numerous studies.\cite{breynaert2020water,hummer2001water,algara2015square,chaban2012confinement,secchi2016massive,algara2015square,sugahara2019negative,chaban2012confinement,liu2015graphene}.\\

It is well appreciated that water near interfaces is perturbed both in terms of its structural and dynamical properties. In the case of hydrophilic surfaces, water forms interactions with the interface and subsequently leads to a slow-down in the water dynamics by a factor of about 4-7. On the other hand, near hydrophobic interfaces, there is a length scale dependence to the dewetting behavior\cite{chandler2005interfaces}. As one brings two interfaces close to each other, we enter the regime of confined water. There have been many experimental and theoretical studies investigating how both thermodynamic and dynamical properties of water change under confinement\cite{ballenegger2005dielectric,verdaguer2006molecular,yuet2010molecular,li2012wetting,sharma2005intermolecular,bonthuis2011dielectric,ballenegger2005dielectric,cicero2008water,koga2001formation,werder2003water,bonthuis2012profile,varghese2019effect,kornyshev1982nonlocal,fumagalli2018anomalously}. It is beyond the scope of the current paper to review all this literature, however we note that there are several interesting thermodynamic and dynamic anomalies that have been observed when water is placed in confined conditions. In this work, we focus on how dielectric properties of water change when sandwiched between two graphene sheets.
	
In previous reports, it was shown that the graphene-water interface induces an orientational polarization of water in close proximity to the surface.\cite{ballenegger2005dielectric,verdaguer2006molecular,yuet2010molecular,li2012wetting,sharma2005intermolecular,bonthuis2011dielectric,ballenegger2005dielectric,cicero2008water,koga2001formation,werder2003water,bonthuis2012profile,varghese2019effect,kornyshev1982nonlocal,fumagalli2018anomalously} This orientational configuration directly affects other properties of water at the interface such as the dielectric response.\cite{ballenegger2005dielectric,verdaguer2006molecular,yuet2010molecular,li2012wetting,sharma2005intermolecular,bonthuis2011dielectric,ballenegger2005dielectric,werder2003water,bonthuis2012profile,varghese2019effect,kornyshev1982nonlocal} However, there have been relatively fewer studies investigating how the dielectric constant of water changes under confinement using both experiments and simulations.

Recently, Fumagalli and co-workers \cite{fumagalli2018anomalously} have used AFM methods to study the molecular polarization of water molecules by applying AC voltage between AFM tip and the surface of graphite. The results of this experiment confirm the suppression of the polarizability of confined water within nano capillaries. Because of the possible arrangement of interfacial molecules and bulk water, there are different charge distributions in these two regimes resulting in a surface capacitance. This feature has been used in order to measure the perpendicular dielectric constant of water under various confinements. It is known that the perpendicular component of the dielectric constant is related to the zeta potential and surface capacitance. \cite{fumagalli2018anomalously,bonthuis2012unraveling,schlaich2016water,konatham2013simulation}.

One of the features of water that makes it unique is its large polarizability. This gives bulk water its large static dielectric constant. Several previous studies have examined how the dielectric properties of water change under confinement.\cite{ruiz2020quantifying,bonthuis2012profile} Water confined between two graphene walls exhibits anomalous behavior because of the suppression of polarization. This results in drastically low values of perpendicular dielectric response.\cite{bonthuis2011dielectric}. In addition, it is worth noting that the structuring of water near interfaces in general is thought to be the dominant factor in controling the position-dependent dielectric permittivity of the system.\cite{varghese2019effect}

The motivation for the calculations reported in this work come from some recent experiments measuring the anomalous dielectric properties of water under confinement\cite{fumagalli2018anomalously}. Specifically, in this work by using a combination of height tunable method for two dimensional capillaries made by atomically flat walls and scanning dielectric microscopy through electrostatic force detection by means of atomic force microscopy (AFM), the molecular polarization of confined water has been measured. In this experiment, by applying a low-frequency ac voltage between the AFM tip and the bottom of graphite electrode, tip-substrate can be identified as first derivative of local capacitance $\frac{dC}{dz}$ in vertical direction. The measured data indicated that epsilon and consequently polarizability of the confined water intensely suppressed in nano-channels with less than $10\;nm$ height.

In this work, the dielectric behavior of water between two graphene layers has been determined via molecular dynamics (MD) simulations. We show that the effective perpendicular dielectric constant of water reduces substantially when the distance between two graphene layers decreases below $\sim 2 $ nm in complete agreement with recent experimental results.\cite{fumagalli2018anomalously} We have also investigated the local value of dielectric permittivity of water as a function of different extents of confinement and find that it reduces to near vacuum permittivity for regions close to the surface. This has also been the subject of some recent simulation work\cite{} although the molecular signatures in terms of both static and dynamical properties have not been fully investigated\cite{jalali2020out,calero2020water,varghese2019effect}. To understand better the origins of the changes of the dielectric properties under confinement, we determine some structural and dynamical properties of water. Specifically, we compute the residence times and diffusion constant of water as a function of confinement. We find that water molecules at the early layers from the surface tends to keep their location for longer amount of time than when they are in the bulk regions. The residence time as well as  diffusion constant results are consistent with the presence of a layer of structured water near the surface of graphene.\\ 
	
The manuscript is organized in the following manner. We begin in Section~\ref{sec:computationalmethods} by introducing the computational method we have used in this work. Then in Section~\ref{sec:dielectric}, we present the results on dielectric properties of confined water between graphene walls as a function of distance from the surface as well as the effective values dielectric constant for various thicknesses of water slabs. Additionally, we relate the permittivity results to the other properties of water such as its hydrogen bond network. In Section~\ref{sec:Dynamics}, we present the residence time as well as diffusion constant as dynamical analysis in order to make a coherent picture of structure and dynamics in surface of water under confined condition.
	
\section{Computational Methods}	\label{sec:computationalmethods}

We use the SPC/E model of water \cite{berendsen1987missing} for our simulations, which has been shown to reproduce the dielectric properties of water rather well. The density of water is $1$ g/ml in slab geometries, ranging from 534 (a monolayer of water) up to 184206 molecules for nearly bulk conditions, which spans different layered structures of confined water molecules. In the version $5.2$ of the GROMACS package\cite{van2005gromacs} we have used the force field proposed by Werder et. al \cite{werder2003water} for water-carbon interactions. This potential has previously been validated by reproducing the contact angle of water near graphene.\cite{werder2003water} The temperature of the system for all simulations is 298.15 K using the Nose-Hoover thermostat.\cite{nose1984unified,hoover1985canonical} The graphene walls consist of carbon atoms with zero charge and are completely flat and parallel to each other. Periodic boundary conditions were applied in all three directions. The box dimensions are: $6.63nm\times6.95nm\times (L_{z}+2nm)$ where  $L_{z}$ is the distance between graphene walls between which the water is placed. \\

\begin{figure}
		\centering
		\includegraphics[width=8.8 cm]{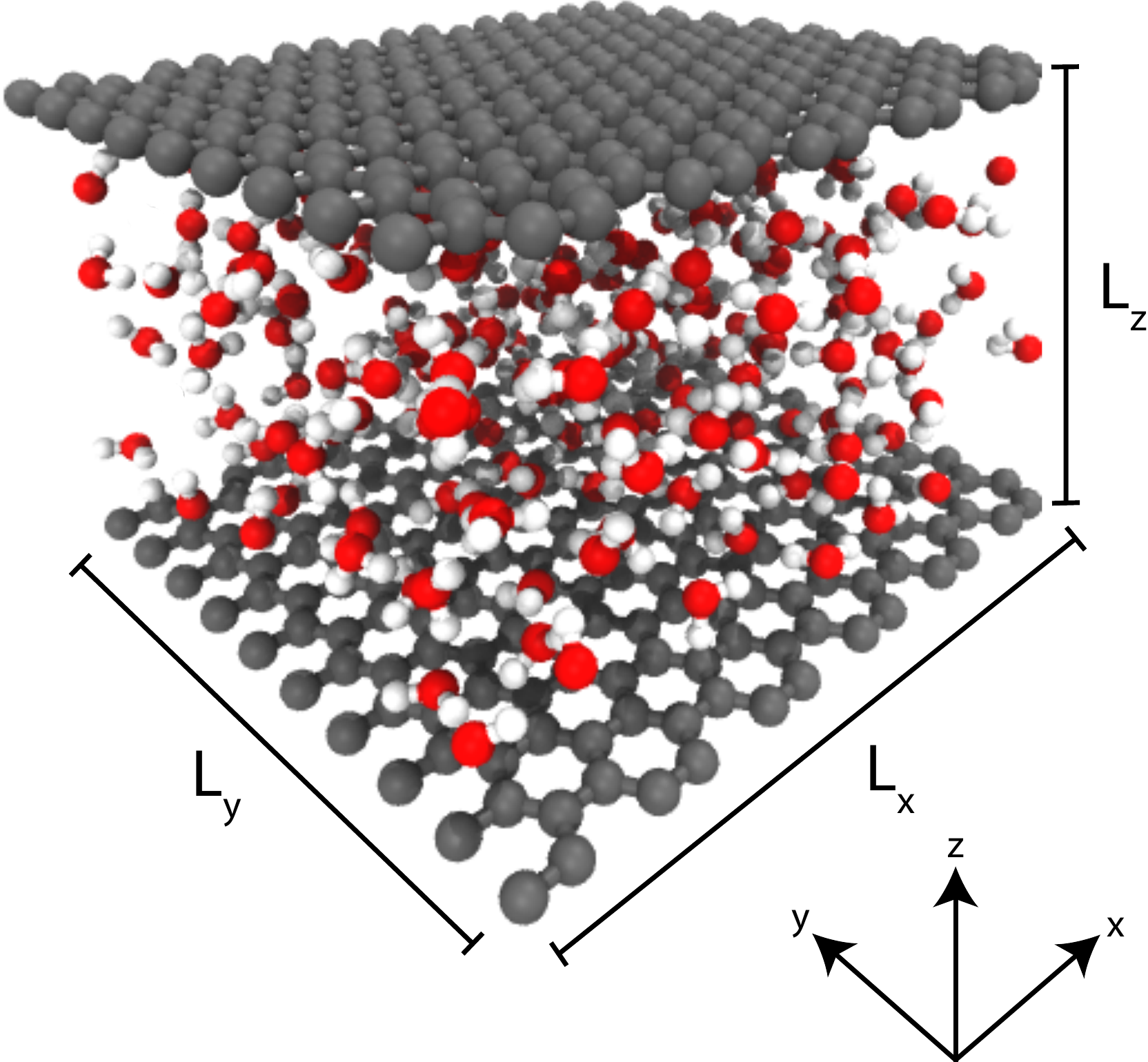}
		\caption{Water molecules under two dimensional slits and graphene walls. Grey balls represent carbon atoms, oxygen as red and hydrogen as white balls. This figure is not scaled in the $ x $ and $ y $ directions according to real systems cell parameters.}
		\label{fig:general-view}
	\end{figure}

Simulations were run for $ 0.2 \; \mu s $ for each distance below $ 100\AA $ and $ 10 \; ns $ for thicker height of two-dimensional slits, using $ 1\text{ fs} $ time step. The truncated radius for the Lennard-Jones interactions is $ r = 1.4 \; nm $ using the Verlet scheme and particle mesh Ewald (PME) summation\cite{essmann1995smooth} is used for treating the electrostatic interactions. A schematic view of the graphene-water channel system is shown in Fig.~\ref{fig:general-view}.
	
\section{Results}
	
\subsection{Dielectric Response}\label{sec:dielectric}

In this section, we present our analysis on the perpendicular dielectric response of water and how it changes as a function of distance from the surface. The change in dielectric displacement field is related linearly to the change of the electric field in the linear response regime. In the Appendix~\ref{app:reviewtheory}, the linear response theory of dielectric constant has been reviewed \cite{bonthuis2012profile,kornyshev1982nonlocal,schlaich2016water,bonthuis2011dielectric,ballenegger2005dielectric}. The most relevant equation for the dielectric response across the graphene-water channel that is used in our simulations is the following:

\begin{equation}
\label{eq:perpepsz}
\varepsilon^{-1}_{\perp}(z) = 1 - \frac{\Delta m_{\perp}(z)}{\varepsilon_{0}k_{B}T + C_{\perp}/V},
\end{equation}
where $\Delta m_{\perp}(z)$ is the polarization along $z$ direction, $C_{\perp}$ is the polarization correlation function integrated over height $z$ as calculated in Appendix~\ref{app:reviewtheory}. $V$ is the volume of the water slab. $T$, $k_{B}$, and $\varepsilon_{0}$ are temperature, Boltzmann's constant, and vacuum permittivity, respectively.\\
	
\begin{figure}[!htbp]
		\centering
		\includegraphics[width=1.0\linewidth]{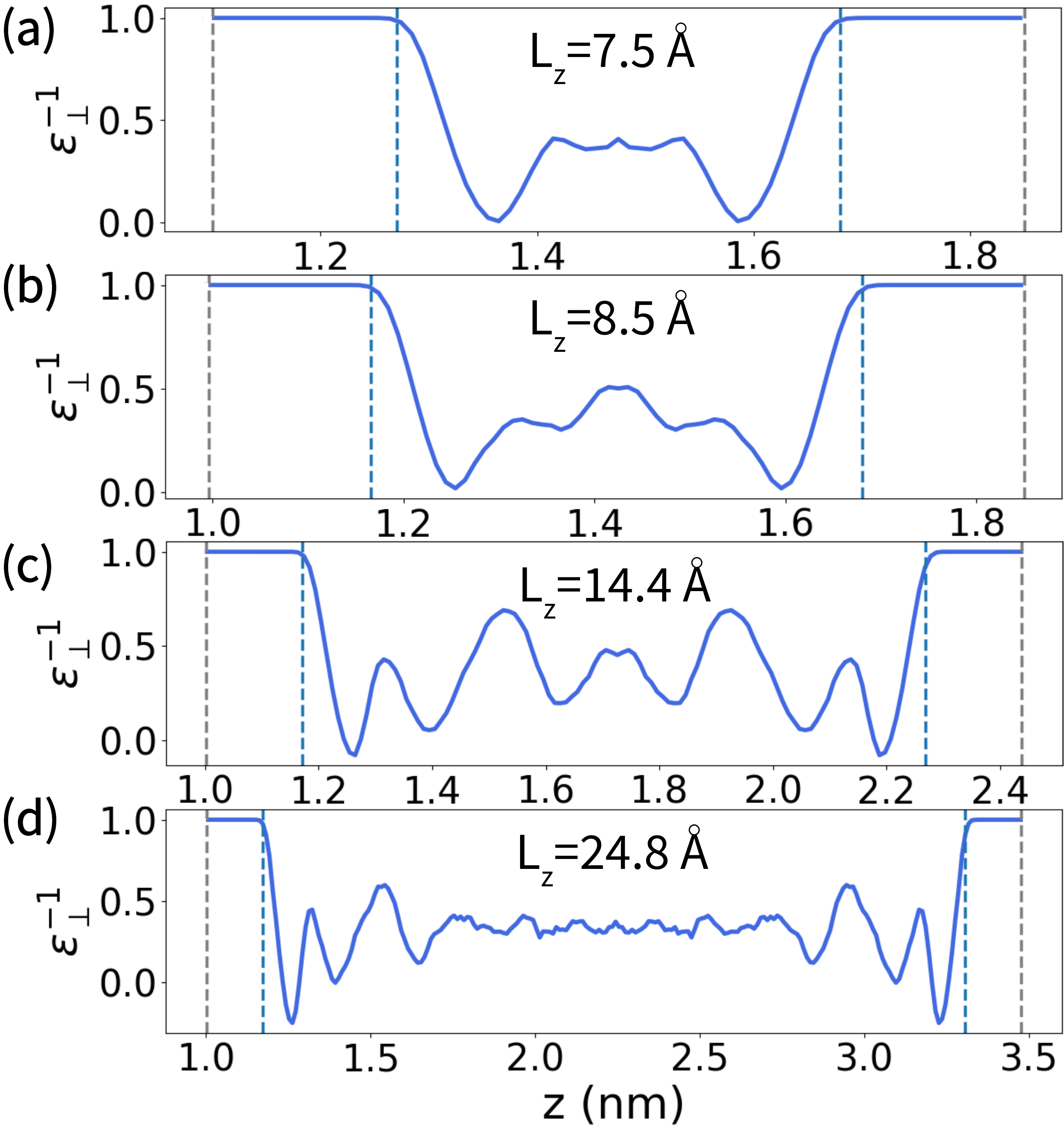}
		\caption{ %
			Inverse of perpendicular component of dielectric constant for different walls distances as a function of height z. Gray and blue vertical dashed lines indicate the surfaces of graphene and water, respectively. Different thicknesses of water slabs are as (a) $ L_{z} = 7.5\AA $, (b) $ L_{z} = 8.5\AA $, (c) $ L_{z} = 14.4\AA $, and (d) $ L_{z} = 24.8\AA $.}
		\label{fig:eps}
\end{figure}

The calculated results for the inverse of the perpendicular component of dielectric constant $ \varepsilon_{\perp} $ as a function of inter-distance of graphene walls for different water slab are presented in Fig.~\ref{fig:eps}. In close proximity to the graphene surfaces, the dielectric function of water in the perpendicular direction is significantly reduced, consistent with previous studies \cite{werder2003water,bonthuis2012profile,ballenegger2005dielectric,hasted1973aqueous,sharma2007dipolar}. Interestingly, under confinement there are some marked oscillations in the perpendicular component of the dielectric constant that acquire even negative values near the interface. Generally, the negative values are due to the external field overscreening of interfacial water molecules. Therefore, water dipolar high polarization induces an inverse electric field, resulting in negative values of dielectric function.\cite{bonthuis2012profile,ruiz2020quantifying,schaaf2016spatially,bopp1998frequency} We will discuss more about these oscillations and relate them to other properties of water in subsequent sections.

\begin{figure}[!htbp]
	\centering
	\includegraphics[width=1.0\linewidth]{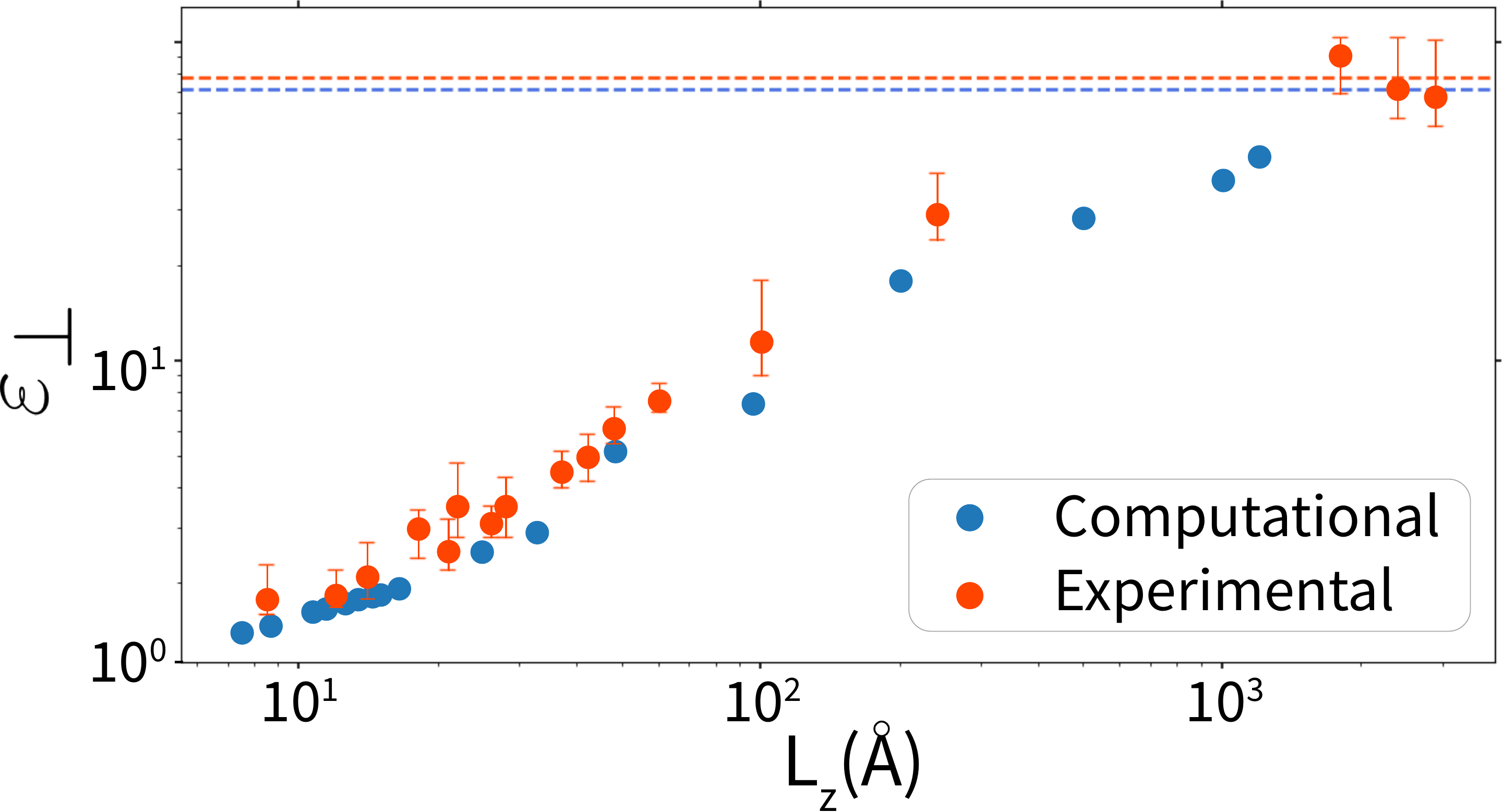}
	\caption{ %
	Computational (blue data) and experimental (red data) perpendicular component of dielectric constant for different distances of graphene walls. Corresponding Dashed lines indicate the bulk values of the SPC/E water model ($\sim 71$) and pure water. The error bars of computational data are small, staying inside of each data circle.
	}
	\label{fig:perpendicular}
\end{figure}

In order to relate the perpendicular component of dielectric constant shown earlier, to the experimentally measured \emph{effective} dielectric constant\cite{fumagalli2018anomalously}, some preliminary steps are needed. Integrating $ \varepsilon_{\perp}^{-1} $ in Eq.\ref{eq:perpepsz} over the channel, results in Eq.\ref{eq:effective-epsilon} which has two unknowns  $ L_{\perp}^{eff} $ and $ \varepsilon_{\perp}^{eff} $. Since we have two of these unknown parameters, we decided to determine the value of $L_{\perp}^{eff}$ by first using the value of $ \varepsilon_{\perp}^{eff} $ for the smallest confined system and using that to reverse engineer the value of $L_{\perp}^{eff}$. Specifically, we used the initial value $ \varepsilon_{\perp}^{eff} = 1.4 $ for the $L_{z}=7.5\AA$ experimentally measured as the narrowest inter-distance between graphene walls, and obtained $L_{eff}-L_{z}=0.94\AA$.\cite{schlaich2016water} Assuming that the difference between $ L_{\perp}^{eff} $ and $L_{z}$ remains constant for higher $L_{z}$ and therefore, by knowing $L_{eff}$, we can determine the $\varepsilon_{\perp}^{eff}$ for all $L_{z}$. These results are shown in Fig.~\ref{fig:perpendicular}. For clarity, the reader is reminded that in Eq.~\ref{eq:effective-epsilon}, $L_{w}$ corresponds to the length of the region where the density of water is non-zero (distance bbetween two blue-dashed lines in Fig.~\ref{fig:density-m-e} in Appendix.~\ref{app:complementary}).

\begin{equation}
	\label{eq:effective-epsilon}
	\int_{-L_{w}/2}^{L_{w}/2} \varepsilon_{\perp}^{-1}(z) dz =  L_{\perp}^{eff}( \frac{1}{\varepsilon_{\perp}^{eff}} - 1 ) + L_{w}.
\end{equation}

Using Eq.~\ref{eq:effective-epsilon}, $L_{eff}$ can be calculated which offers from $L_{w}$. Assuming same values for the difference between acquired $ L_{\perp}^{eff} $ and $ L_{w} $ for other confinements, we can calculate the $ \varepsilon_{\perp}^{eff} $ for all other water slab thicknesses. As we can observe from Fig.~\ref{fig:perpendicular}, the perpendicular component is very low and close to $ 1.4 $ in all slab thicknesses less than $ \sim 15\AA $ and then increase to bulk values gradually. The trend of our calculated results are in good agreement with recent reported experimental data \cite{fumagalli2018anomalously}. Interestingly, the length scale over which it takes for the dielectric constant to converge to the bulk value is quite large (over 1000 \AA) and is much longer than that needed to converge radial and orientational correlations.

\begin{figure}[!htbp]
		\centering
		\includegraphics[width=1.0\linewidth]{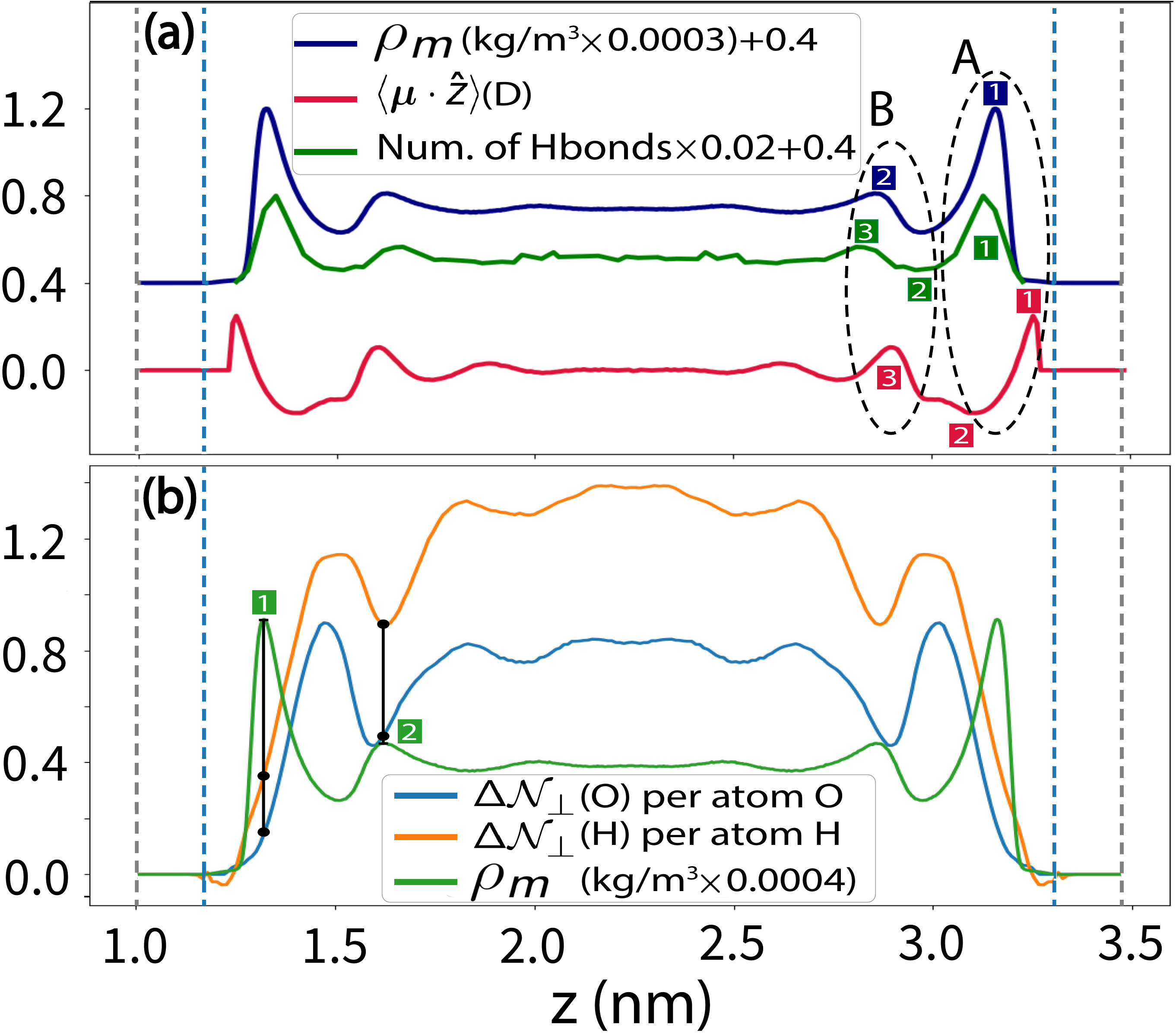}
		\caption{ %
			(a) The time-averaged $z$ component of dipole moments of water molecules as a function of height $z$. Regions $A$ and $B$ correspond to the first and the second layer estimated by density of particles (blue curve), respectfully. Additionally, the green curve presents the number of Hbonds. The slab thickness is $L_{z} = 24.8\;\AA $. The particle density and number of hydrogen bonds have been multiplied by $0.0003$ and $\frac{1}{50}$, respectively, as a rescaling for a better display.(b) The fluctuations in the number of hydrogen and oxygen per atom of each type as a function of height z for thickness $ L_{z} = 24.8\AA $. The diagrams of $ \Delta \mathcal{N}_{\perp}(H) $, $ \Delta \mathcal{N}_{\perp}(O) $, and density of particles have been multiplied by 20, 20, and $ \frac{1}{2500} $. }
		\label{fig:DenO-DeltaRhoO-DeltaRhoH}
	\end{figure}

\begin{figure}[!htbp]
		\centering
		\includegraphics[width=1.0\linewidth]{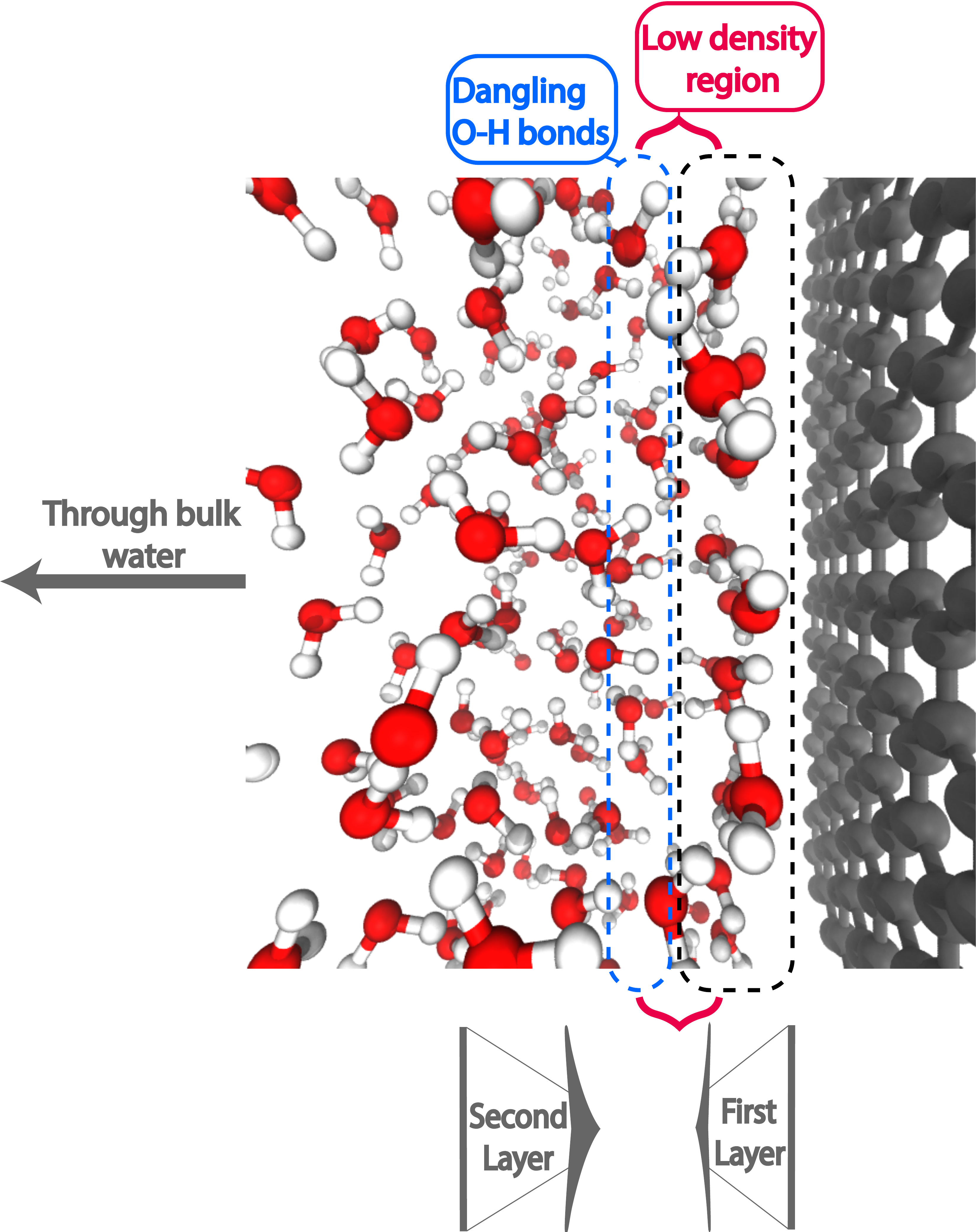}
		\caption{%
			Schematic figure for a few layers of water near the interface of water-graphene.
		}
		\label{fig:schematic}
	\end{figure}

We next turn to examining the evolution of the hydrogen bond network in order to understand why the dielectric properties change under confinement. We begin first with understanding how the orientation of water changes near the graphene layer. In order to analyze the orientation of hydrogen bonds as a function of $z$ in the slab of water, we calculated the time-averaged $z$ component of dipole moments for water molecules. The maximum $1$-red in Fig.~\ref{fig:DenO-DeltaRhoO-DeltaRhoH}(a) adjacent to the graphene walls, demonstrate the tendency of the O-H bonds to interact with carbon atoms and as a result, be directed into the graphene sheets. In the first interfacial layer shown by oval-dashed line $A$, due to the hydrogen bond network, most of the water molecules spend time in a planar configuration (see also Fig.~\ref{fig:schematic}). However, there is a small tilting from the $x-y$ plane by $\sim 10^{\circ}$ corresponding to the minimum $2$-red in Fig.~\ref{fig:DenO-DeltaRhoO-DeltaRhoH}(a). This can be considered as the criteria of the planar hydrogen bond for water molecules at the interface\cite{varghese2019effect}. 

Since the sign of averaged dipole moment in the minimum $2$-red is opposite to the maximum $1$-red, we conclude that the deviation from planar configuration at the bottom of the first layer is directed into the low-density region below the first layer. The non-zero value of dipole moment in $z$ direction in the second layer, corresponding to oval-dashed line $B$, (maximum $3$-red and $2$-blue) is consistent with the dangling O-H bonds as discussed in the previous sections. As it is shown in the region $B$ in Fig.~\ref{fig:DenO-DeltaRhoO-DeltaRhoH}(a), the $z$ component dipole moment in the maximum 3-red has the same sign as the hydrogen bonds adjacent to the graphene walls, meaning that dangling O-H bonds are directed into the low-density region between the first and the second layers of interfacial water (see also Fig.~\ref{fig:schematic}). To further assert this argument, the number of Hbonds have been computed as a function height $z$ (Fig.~\ref{fig:DenO-DeltaRhoO-DeltaRhoH}(a)). The hydrogen bond criterion was determined by the standard Luzar and Chandler recipe\cite{luzar1996hydrogen}. The maximum $1$-green indicates the high number of Hbonds in the first layer as we expect from Hbond network. The minimum $2$-green in Fig.~\ref{fig:DenO-DeltaRhoO-DeltaRhoH}(a) corresponds to the lack of Hbonds in the low-density region and early parts of second layer. The minimum $2$-green has entered into to second layer (maximum $2$-blue) that indicates water molecules at early parts of second layer create less Hbonds in comparison to the deeper parts of water that indicates the presence of dangling O-H bonds. 

The preceding analysis does not inform on the changes in the fluctuations of the various degrees of freedom involving the solvent as it approaches the graphene layers. To this end, we computed the Fano factors associated with the oxygen and hydrogen atoms individually as a function of position along $z$ direction (see  Fig.~\ref{fig:DenO-DeltaRhoO-DeltaRhoH}(b)). The Fano factor is defined as:
	
\begin{equation}
\Delta \mathcal{N}_{\perp,X}(z) \equiv \frac{\langle N_{X}^{2}(z)\rangle - \langle N_{X}(z)\rangle^{2}}{\langle N_{X}(z)\rangle},
\end{equation}
	
Note that $\langle ... \rangle$ is the time average over the whole trajectory and $N_{X}(z)$ is the number of atom $X$ (oxygen or hydrogen) in height $z$. It is worth mentioning that several previous studies have shown that the Fano factor can be used to probe the local compressibility of water\cite{}. As expected, due to the lighter mass of hydrogen it is characterized by larger Fano factors. In the first layer (maximum $1$ in Fig.~\ref{fig:DenO-DeltaRhoO-DeltaRhoH}(b)), the ratio of fluctuations in number of hydrogen atoms to fluctuations in number of oxygen atoms is $ \frac{\Delta \mathcal{N}_{\perp,H}}{\Delta \mathcal{N}_{\perp,O}} \simeq \frac{0.35}{0.15} = 2.33 $, while this ratio for the second layer(maximum $2$ in Fig.~\ref{fig:DenO-DeltaRhoO-DeltaRhoH}(b)) is $ \frac{\Delta \mathcal{N}_{\perp,H}}{\Delta \mathcal{N}_{\perp,O}} \simeq \frac{0.9}{0.5} = 1.8 $. This indicates that although fluctuations in the number of both hydrogen and oxygen atoms in the first layer are smaller than the second layer and the rest of water, the ratio of fluctuations for the first layer is larger than the remaining part of water. Consistent with the results presented in the previous sections, in the first layer, hydrogen atoms can rotate around oxygens (as mostly fixed points) in order to align the water molecule along the direction of the applied field. However, this behavior is reversed in bulk water, where the whole water molecule tends to rotate around a point on the dipole vector.
	
The molecular signatures of water's hydrogen bond network near the graphene surface has profound implications on the dielectric constant and how it evolves as a function of distance from the graphene surface. Consistent with recent studies\cite{varghese2019effect} the most energetically favorable and therefore most probable orientations of water molecules close to graphene tend to have their dipole oriented parallel to the solid surface and creating a network of Hbonds at the interface of graphene-water. This hydrogen bond network cannot be aligned along the perpendicular directed external field for the response, and as a result, creates an over screening layer for the external electric field. 
	
\begin{figure}[!htbp]
		\centering
		\includegraphics[width=1.0\linewidth]{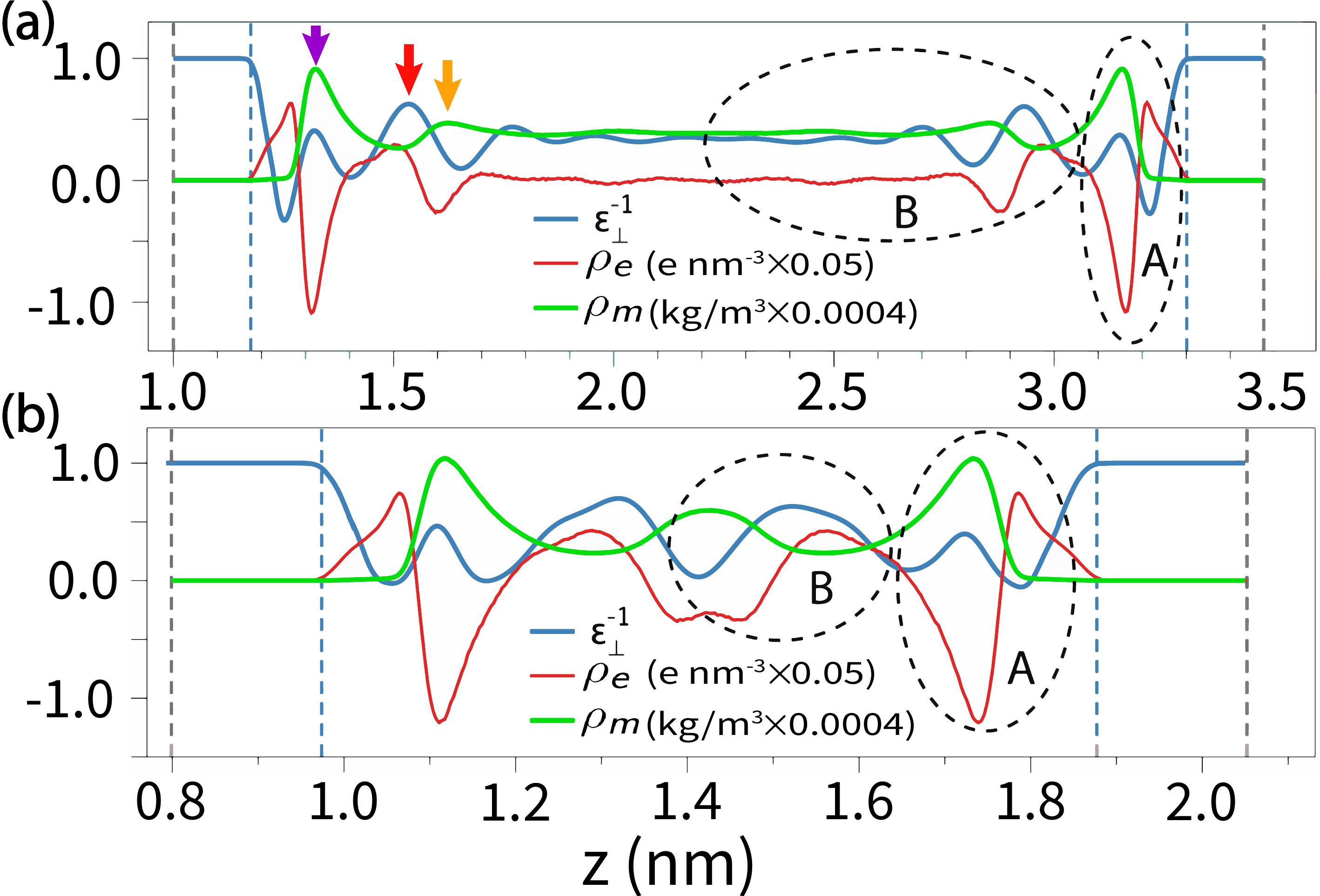}
		\caption{ %
			(a) The purple marker and $\varepsilon_{\perp}^{-1}$ maximum determine the restriction of water molecules in order to rotate out of the $x-y$ plan, causing very low dielectric constant. The red area is compatible with the low-density region between the first and second layer of the water molecules. The orange zone showing the presence of dangling OH bonds, leading to higher values for the $\varepsilon_{\perp}$. The slab thickness is $L_{z} = 24.8\;\AA $.(b) Inverse of the perpendicular component of dielectric constant versus density of charges as well as density of particles. Regions A show the opposite behavior of $ \varepsilon_{\perp}^{-1} $ and density of charges $(\rho_e)$. Regions B indicate the consonant trend between $ \varepsilon_{\perp}^{-1} $ and $\rho_e$. The water thickness is $L_{z} = 12.5\;\AA $. The diagrams for $\rho_e$ and $\rho_m$ have been scaled up via dividing the values by 20 and 2500, respectively.}
		\label{fig:corr-charge-eps}
	\end{figure}

As can be seen earlier from Fig.~\ref{fig:eps}, there are a set of maximums and minimums close to the interface of water. The first maximum in $ \varepsilon_{\perp}^{-1}(z) $ is related to the first layer of water (area as indicated by purple marker in Fig.~\ref{fig:corr-charge-eps}(a)). This maximum can be assigned to a network of Hbonds in water molecules mostly parallel to the graphene sheets. This layer is responsible for screening effects and interrupting this network via an external field is energetically unfavorable. Another consequence of this screening effect is the diverging or more intensively the negative values for $ \varepsilon_{\perp}(z) $ (the first minimums at interfaces in Fig.~\ref{fig:eps} and Fig.~\ref{fig:corr-charge-eps} for $ \varepsilon^{-1}_{\perp} $) close to the surface of water. 

In order to dig deeper into the origins of the oscillations of the dielectric constant we investigated the coupling between the mass and charge density of particles as shown in Fig.~\ref{fig:density-m-e}. We begin with understanding the evolution in the mass density. Due to the separation of the first layer with the rest of water molecules, there is a low-density region as thin as $ 2\AA $ between the first and the second layers, which is mostly filled with hydrogen atoms rather than oxygen ones. This region is shown schematically in Fig.~\ref{fig:schematic} in the yellow region. Because of the lack of water molecules in this region in order to respond to the external electric field, $\varepsilon_{\perp}^{-1}$ reaches its maximum value in comparison to the other maximums in the graph (red area in Fig.~\ref{fig:corr-charge-eps}(a)). It can be concluded that water molecules in the second layer cannot make significant Hbonds with the first layer. As a result, they are left with dangling OH bonds as shown in Fig.~\ref{fig:schematic}. Since these dangling OH bonds have more freedom to rotate especially out of the $x-y$ plan, they can respond to the external field stronger than OH bonds in the first layer which are involved in a robust hydrogen bond network. This higher response leads to higher values of $\varepsilon_{\perp}(z)$. A few more tenths of Angstrom further into the bulk after the second maximum in the density of particles, $\varepsilon_{\perp}$ reaches a maximum. This is in agreement with the existence of dangling OH bonds in the second layer (orange area in Fig.~\ref{fig:corr-charge-eps}(a) and blue dashed rectangle in Fig.~\ref{fig:schematic}). Note that this minimum in $\varepsilon_{\perp}^{-1}$ occurs for all thicknesses of water slab considered in our work and thus appears to be a generic phenomenon that occurs not only for water under confinement.
	
Besides the mass density, there are also some interesting features involving the coupling between the charge density and dielectric constant. In the proximity of the graphene sheets, the changes of the charge density and the $\varepsilon_{\perp}^{-1}$ are anticorrelated while in the bulk, they are correlated (see Fig.~\ref{fig:corr-charge-eps}, marked as A and B, respectively). This indicates that in the first layer, oxygen atoms can hardly participate in dielectric response due to their heavier mass and rigidity in movement in the hydrogen bond network as explained earlier. 

A simple picture of configuration is that each oxygen atom can host two Hbonds in addition to the two covalent bonds with the hydrogen atoms within the water molecule. Thus, oxygen atoms are involved in creation of four Hbonds. In contrast, hydrogen atoms can only be involved between two oxygen atoms in a Hbond. Hence, out-of-plane movement of oxygen atoms can destruct the Hbond network more effectively than hydrogens' out-of-plan rotation. Therefore, the rotation of hydrogen toward graphene layer around oxygen atoms is the dominant response mechanism to the external field in region A. In contrary, in the second layer as well as the rest of water, the changes in the charge density and $ \varepsilon_{\perp}^{-1} $ are consonant. This reflects that rotation of both oxygen and hydrogen atoms together around the center of mass of the water molecule contribute as the response mechanism. Due to the absence of Hbond network here, this rotation can occur without energetically unfavorable disturbing any Hbond network. Therefore, since the center of mass of water molecule is closer to the oxygen atom than hydrogen atom, in comparison with charge density the response to the external field becomes consonant with the presence of oxygen atoms (Fig.~\ref{fig:corr-charge-eps}).

\subsection{Dynamical Properties} \label{sec:Dynamics}
	
In the previous section, we have focused on identifying static molecular signatures of the hydrogen bond network in order to correlate them with the behavior in the dielectric constant. We move next to understanding and searching for whether these features are also reflected in the dynamic quantities. Specifically, we focus on two dynamical properties namely the residence time and diffusivity. We have followed the residence time of water molecules for different layers of confined water . The definition of residence time correlation function is given by the following equation as done in numerous previous studies\cite{pizzitutti2007protein, qaisrani2019structural,rana2013ab}.
	
\begin{equation}
	N_{w}(t) = \frac{1}{N_{t}}\sum_{n=1}^{N_{t}}\sum_{}^{i} P_{i}(t_{n},t),
\end{equation}

in which the conditional probability $ P_{i}(t_{n},t) $ is $1$ if the $i$th water molecule remains in the selected region in $ t_{n} $ and $ t_{n} + t $ time interval, otherwise is zero. $ N_{t} $ is the number of time-frames with the length $ t $ that we calculate the survival probabilities. Here, we have estimated different selected layers of water based on the density profiles. Typically, the first layer of water calculated from one of graphene sheets relates to the first maximum of particle density diagram. We have considered $ \sim 3 \AA $ as the thickness of a single layer of water, in which the maximum value of the particle density is in the center of the selected region for each layer. As a result of planar hydrogen bond networks, we expect that the layered structure of water mostly exists at the interface. Therefore, we have selected the first, second, and other possible layers of water slab, and compared them with a segment that has the same thickness in the middle part of the confined system. We also compared these results with a similar segment in pure SPC/E water without confinement.

\begin{figure}[!htbp]
		\centering
		\includegraphics[width=1.0\linewidth]{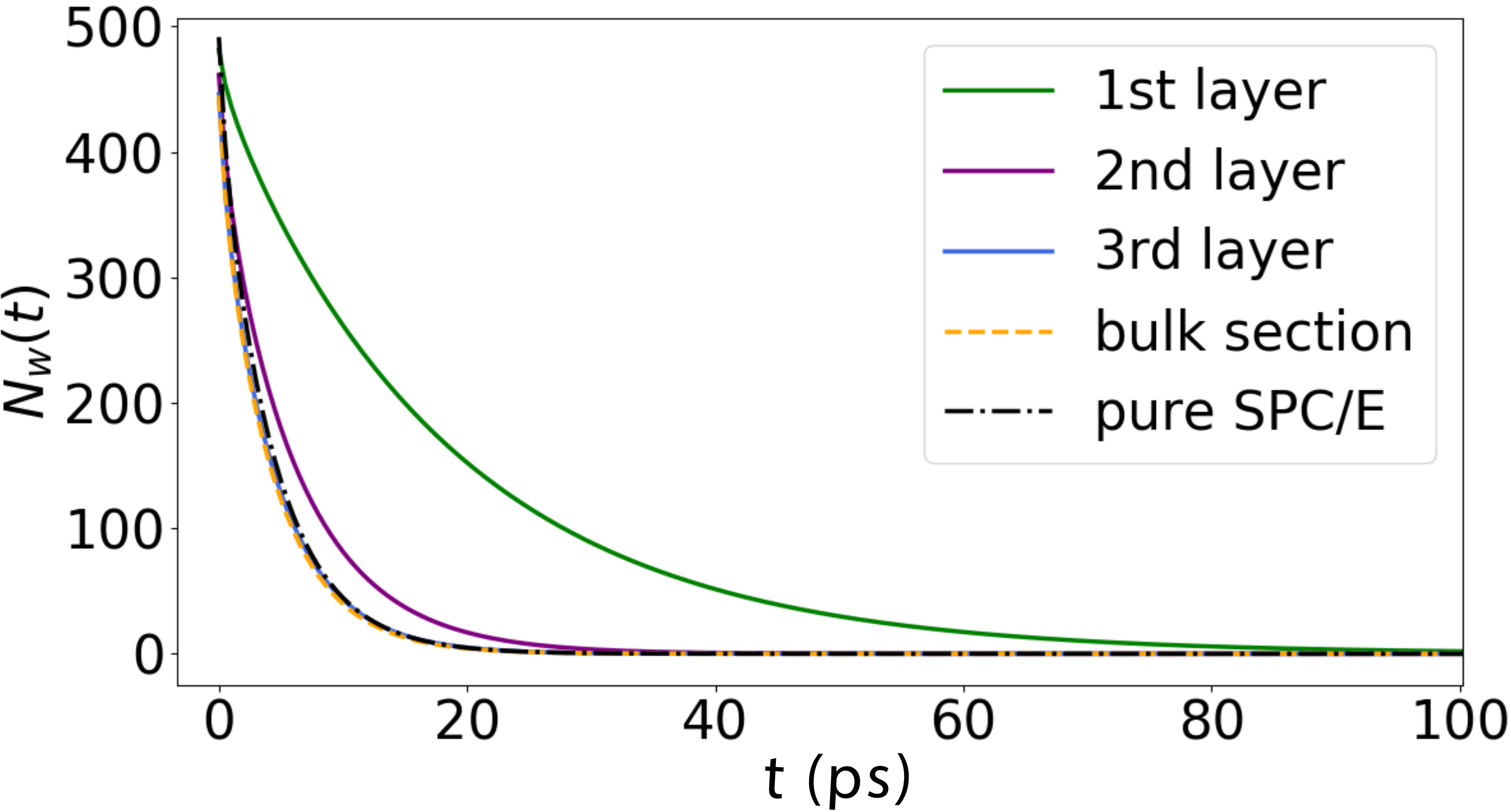}
		\caption{%
			Residence time correlation function of water molecules between graphene walls for $ L_{z} = 24.8\AA $. The residence time is obtained for early layers of each system, indicating that the first layer (green line) has the longest residence time, then the second layer (purple line) is the layer with the highest survival time. The third (blue line) and the bulk segments (orange dashed line) of the confined channel has mostly the same residence time with pure SPC/E water (black dash-dot lines) without confinement.
		}
		\label{fig:RTs}
	\end{figure}

As can be seen from Fig.~\ref{fig:RTs}, the residence time of confined water molecules at the interface (green and purple lines) is more than the residence time of molecules in the bulk (dashed yellow and dashed-dot black lines).  This indicates that the strength of Hbonds in the first layer's hydrogen network is higher than areas closer to the bulk and water molecules tend to stay at this layer for a longer period of time. This observation is in agreement with our earlier observations of the existence of strong Hbonds network in this layer. 
	
	In order to obtain more quantitative measures on the timescales associated with water exchange in the different layers near graphene and how it compares to the bulk, we found that they could be fit to a maximum of two exponential functions of the following form: 
	
\begin{equation}
	N_{w}(t) =  n_{1}e^{-t/\tau_{1}} + n_{2}e^{-t/\tau_{2}} + n_{p}.
	\label{eq:RTfit}
\end{equation}

Here, $ n_{p} $ is the number of water molecules which permanently and continuously are in the selected region. We have summarized the data for the system with $ L_{z} = 24.8\AA $ in Table.~\ref{tab:RT-248A}.
	
\begin{table*}[!ht]
	\caption{Residence time results for \( L_{z} =24.8 \AA \) distance between graphene sheets.}
	\centering
	\begin{tabular}{ l  |  l   c   c   c   c   c   c   c }
		\hline
		\hline
		\( \text{Estimated Layers} \) & \( \tau_{1}(ps) \) & \( \tau_{2}(ps) \) & \( n_{1} \) & \( n_{2} \) & \( n_{p} \)  \\
		\hline
		1st layer & 18.44 & 1.18 & 450.61 & 23.88 &  \(\sim 0\)\\
		2nd layer & 6.34 & 0.97 & 395.16 & 53.22 & \(\sim 0\) \\
		3rd layer & 4.41 & -  & 401.35 & - & \(\sim 0\) \\
		Middle layer (close to bulk) & 4.24 & - & 400.18  & - & \(\sim 0\)\\
		Pure SPC/E water & 4.28 & - & 445.85 & - & \(\sim 0\)\\
		\hline
		\hline	
	\end{tabular}
	\label{tab:RT-248A}
\end{table*}

As can be seen from \ref{tab:RT-248A} (also from \ref{tab:RT-124A}, \ref{tab:RT-164A}, \ref{tab:RT-428A} shown in the Appendix~\ref{app:complementary}), one needs two exponential terms for describing the behavior of residence time for first and the second layers. This can be attributed to two populations of water molecules, one that is involved in a more strong hydrogen bond network and another that is more labile as discussed in the previous section. As one moves from the interface to the bulk, there is a reduction in the residence time by about a factor of 4 which is consistent with the notion that the bulk-region of water is characterized by less strong hydrogen bonds. It should also be stressed that the slow-down in the water dynamics is not so drastic in the sense that the graphene-water interface is still a very dynamic system on the 10s of picosecond timescale. Similar features have also been observed using ab initio molecular dynamics simulations of water near graphene by Chandra et al.\cite{kayal2019water}.

In order to understand better how the translational motion of water molecule change near the graphene surface, we computed the translational diffusivity from the mean square displacement of the oxygen atoms of the water as described by the canonical equation below:

\begin{equation}
	D_{s} = \frac{1}{N}\sum_{n=1}^{N} (r_{n}^{\alpha}(t) - r_{n}^{\alpha}(0))^{2},\;\;\alpha = x,y
\end{equation}

where $ r_{n}^{\alpha} $ is the position of the nth particle, N is the number of particles, and $ t $ is time. The time interval of 500 time steps was chosen in acquiring the slope of MSD versus time that gives the diffusion constant. For our analysis, we found that the linear regime could be fit between XX and YY picoseconds in order to extract the diffusion constant. \\
	
\begin{figure}[!htbp]
	\centering
	\includegraphics[width=1.0\linewidth]{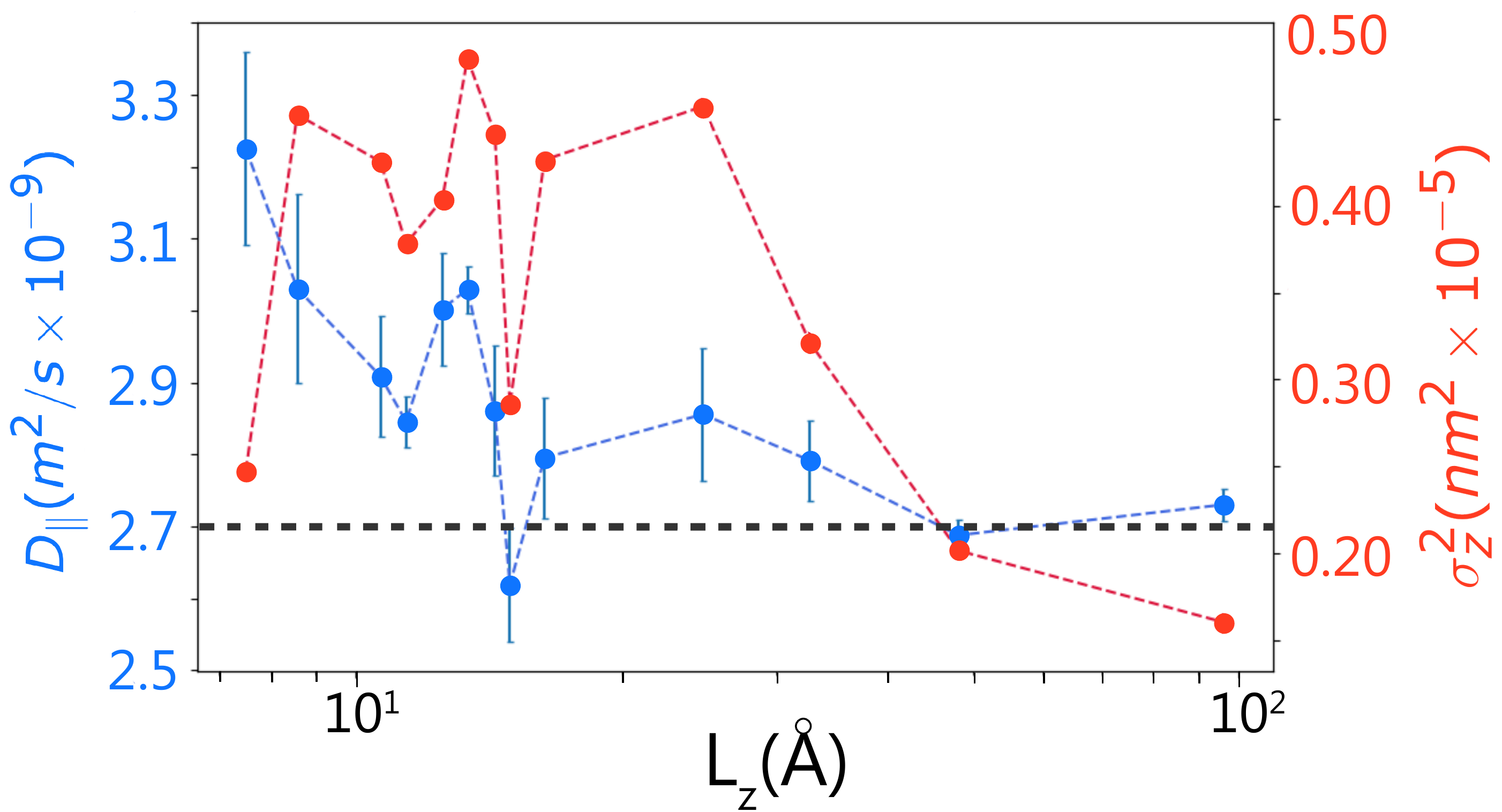}
	\caption{%
		The lateral diffusivity (blue data) and variance in the $z$ component of the center of mass (red data) of confined systems as a function of the thickness of water slab. These two diagrams are shown together to compare the trends. Except for $ L_{z} = 7.5\AA $ which corresponds to one layer of water, the trend of both diagrams for remaining thicknesses is mostly consistent. This shows the effect of commensurability and hopping of water molecules between early layers of water as a factor in determining the lateral diffusivity. The gray dashed line is determining the bulk value of lateral diffusion constant for bulk water without confinement with graphene walls.
	}
	\label{fig:lateralDif_comZfluc}
\end{figure}

Overall, Fig.~\ref{fig:lateralDif_comZfluc} shows that the lateral diffusivity is higher at the interface compared to the bulk. Although the water molecules have a larger residence time at the interface, they are not rigid and exhibit enhanced mobility. This result is also consistent with previous studies\cite{ghorbanfekr2020insights,mark2001structure,cicero2008water} simulating water near graphene surfaces. We propose that the higher diffusivity of water molecules under confinement is due the motion of the entire hydrogen bond network in the horizontal plane and not because of movement of water molecules as single units. This collective movement of water molecules is due to the unfavorable energy that is needed for one single molecule to leave the hydrogen bond network as one would need to break an enthalpically stabilized strong hydrogen bond interaction. Therefore, water molecules tend to keep their collective configuration as planar clusters at the interface during diffusion. The distance that water molecules in a cluster travel is more than the distance of a single water molecule motion in the bulk. This causes the higher lateral diffusion constants for water molecules at the interfaces.\\

In order to understand the underlying mechanism associated with the enhanced lateral diffusion, we computed the time-averaged $z$ component of fluctuation in center of mass of the whole system is calculated in Fig.~\ref{fig:lateralDif_comZfluc} for the different slab thicknesses. The center of mass of the system is removed from the $ \sigma^{2}_{z}$. We found that behavior of variance for confined systems' center of mass along $z$ direction as a function of water slab thickness, mirrors the behavior seen in the lateral diffusion except for the most confined system. Recall that$ L_{z} = 7.5\AA $, corresponds to the thickness of a single layer of water molecules and therefore do not have any space to move in the $z$ direction. However, this single layer of water has the highest lateral diffusion due to the motion of the layer of water. We also note that all the values of $\sigma_{z}^{2}$ for all confined systems are about ten times less than $\sigma_{z}^{2}$ for the bulk water without confinement that is $3.47\times10^{-5}\;nm^{2}$. \\


We can also relate the trends observed in the lateral diffusion to commensurability between the space required for an entire layer of water molecules to be embedded in the confined channel. The effect of commensurability has been observed in the study of other dynamical properties of confined water\cite{neek2016commensurability, gao1997origins,verdaguer2006molecular,cleveland1995probing,jeffery2004direct}. The oscillatory behavior of lateral diffusion in Fig.~\ref{fig:lateralDif_comZfluc} for the channel widths below than $ \sim 30\AA $ is due to the commensurability for early water layers at the interface. This commensurability determines the hopping of water molecules among early surface layers for a slab with a fixed density. Some specific channel sizes, for example with a certain ratio of water molecules in the first and second layer as well as rest of water, cause a minimum fluctuation in the z component in the center of mass of water. These channels are consistent with minimum interlayer water molecules hopping and as discussed previously, minimum lateral diffusion. The same trend is true for other commensurabilities. This oscillatory behavior has been followed experimentally in atomic force microscopy (AFM) experiments\cite{neek2016commensurability, gao1997origins,verdaguer2006molecular,cleveland1995probing}, that the solvation force varies with a period close to water molecule size which is consistent with our computed period of oscillations. Indeed, the AFM results show that the dynamics of confined water depend significantly on the exact confining cavity size which can determine the extent to which the water slab is commensurate with the water molecule thickness is\cite{jeffery2004direct}.
	
Here, we propose that when the number of hopping is larger, therefore there will be more region for the clusters of water molecules in the early layers to move in the parallel direction to the graphene walls. After leaving a water molecule from a primary layer at the interface, adjacent water clusters in that planar layer fill the empty region. As a result, when the number of hopping among the early layers is larger, the lateral diffusion increases. In contrast, when the number of hopping is small, the hydrogen bond networks are mostly compact next to each other and harder to move. This happens, for the channel size of $ L_{z} = 15.4 \AA $, in which the fluctuations in the $z$ component of the center of mass are minimum among confined water slabs with a thickness of less than $ \sim30\AA $. Hence, both the hydrogen bond networks and hopping between layers are deterministic factors to define the behavior of lateral diffusion of water in confined channels.
	

\section{Conclusions}

In this work, we investigated the dielectric permittivity profile of water in slab and confined geometry between two parallel and neutral graphene layers. We considered the profile of \( \varepsilon_{\perp}(z) \) as a function of height z in confined systems and found new relations between this quantity and density of charges as well as fluctuation in the number of hydrogen and oxygen atoms per each atom separately as a function of height z in the channel. These findings lead us to molecular scale differences in types of motion and responses between hydrogen and oxygen atoms at the interface and in the bulk. Also, we follow the details in the changes of the dielectric response and water layers configurations, in which we can locate the dangling Hbonds in the region between the first and second layers of water. \\

Furthermore, we show that the effective perpendicular dielectric constant decreases drastically when the distance between two graphene sheets decreases and reaches to value \( \sim 1.4 \) in most extreme confined systems. The merging of the perpendicular dielectric constant with the bulk value of the SPC/E water model happens for thicknesses that are larger than our expectation of the water slab that should be in the bulk regions. The perpendicular dielectric constants in this work are in agreement with recent experimental data, in which the capacity model for determining the effective dielectric constant has been used among AFM measurements.\\
	
We show, by computing the residence time of water in different layers, that water molecules tend to remain in the first layer more than other parts of the channels, which is consistent with the existence of strong Hbonds in the hydration shell. Therefore, molecules in the second layer cannot make significant Hbonds with the first layer, creating dangling O-H bonds directed into the low-density region. The third layer and middle segment of water slabs have the same residence time in comparison to the pure SPC/E model. This demonstration is in agreement with a small population fluctuation per number of atoms for both hydrogen and oxygen at the interfaces.\\

Finally, we also determined the lateral self-diffusion constant of water under confinement and showed that almost for all channel thicknesses, it is higher than bulk value of the water. Therefore, in spite of the confinements, water molecules can diffuse in collective motions at the interface, causing higher values of lateral diffusivity. The key factor in determining the lateral diffusivity in confined structures for a fixed density, is the number of water molecules hopping among the layers at the interface.

\emph{Acknowledgements}
IA thanks Mehdi Hassanpour for helpful discussion in simulations of this work. AH and IA thank Muhammad Nawaz Qaisrani for the helpful conversation. Authors also acknowledge the HPC center of Sharif University of Technology. 

\appendix

\section{Review on linear response theory of dielectric constant in confined systems} \label{app:reviewtheory}
Here, we review the theory of linear response and fluctuation-dissipation for dielectric function for slab geometries such as our system of study\cite{bonthuis2012profile,kornyshev1982nonlocal,schlaich2016water,bonthuis2011dielectric,ballenegger2005dielectric}. The change in dielectric displacement field is related linearly to the change of electric field in linear response regime via equation
\begin{equation}
	\Delta \boldsymbol{D} = \varepsilon_{0} \int \Delta \boldsymbol{E}(\boldsymbol{r}^{'}) . \varepsilon_{nl}(\boldsymbol{r},\boldsymbol{r}^{'}) d \boldsymbol{r^{'}}
	\label{eq:1}
\end{equation}
where $ \varepsilon_{0} $ is vacuum permittivity, $ E(r^{'}) $ is local electric field and $ \varepsilon_{nl}(\boldsymbol{r},\boldsymbol{r}^{'}) $ is non-local dielectric tensor. A Constant electric field in homogeneous system results in local response function as product of permittivity tensor. Change in electric field is given by

	\begin{equation}
	\Delta \boldsymbol{D} = \varepsilon_{0}\varepsilon(\boldsymbol{r}) \cdot \Delta \boldsymbol{E}
	\label{eq:2}
	\end{equation}
	where $ \varepsilon(\boldsymbol{r}) = \int \varepsilon_{nl}(\boldsymbol{r},\boldsymbol{r}^{'}) d \boldsymbol{r^{'}} $ and using local assumption then we have $ \varepsilon_{nl}(\boldsymbol{r},\boldsymbol{r}^{'}) = \delta(\boldsymbol{r} - \boldsymbol{r}^{'})\varepsilon(\boldsymbol{r}) $.\
	The inverse of dielectric response function is defined similarly
	
	\begin{equation}
	\Delta \boldsymbol{E} = \varepsilon_{0}^{-1} \int \Delta \boldsymbol{D}(\boldsymbol{r}^{'}) . \varepsilon_{nl}^{-1}(\boldsymbol{r},\boldsymbol{r}^{'}) d \boldsymbol{r^{'}}
	\label{eq:3}
	\end{equation}
	where $ \varepsilon_{nl}^{-1}(\boldsymbol{r},\boldsymbol{r}^{'}) $ is the inverse of $ \varepsilon_{nl}(\boldsymbol{r},\boldsymbol{r}^{'}) $ defined by $ \delta(\boldsymbol{r} - \boldsymbol{r}^{'}) = \int \varepsilon_{nl}(\boldsymbol{r},\boldsymbol{r}^{'}) \varepsilon_{nl}^{-1}(\boldsymbol{r}^{'},\boldsymbol{r}^{''})d \boldsymbol{r^{'}} $
	and when the displacement of field is constant, similar to above discussion, the inverse dielectric response function is local and results in\cite{bonthuis2012profile,kornyshev1982nonlocal,schlaich2016water,bonthuis2011dielectric,ballenegger2005dielectric}
	
	\begin{equation}
	\Delta \boldsymbol{E} = \varepsilon_{0}^{-1}\varepsilon^{-1}(\boldsymbol{r}) \cdot \Delta \boldsymbol{D}
	\label{eq:4}
	\end{equation}
	in which $ \varepsilon^{-1}(\boldsymbol{r}) $ is the inverse of dielectric tensor. For the perpendicular component of electric field and dielectric displacement field, we obtain following similar equations \cite{bonthuis2012profile,kornyshev1982nonlocal,schlaich2016water,bonthuis2011dielectric,ballenegger2005dielectric}
	
	\begin{equation}
	\Delta E_{\perp}(z) = \varepsilon_{0}^{-1}\varepsilon^{-1}_{\perp}(z) \Delta D_{\perp}
	\label{eq:5}
	\end{equation}
	
	Electric field is separated into displacement field $ \boldsymbol{D}(\boldsymbol{r}) $, for monopole terms in the integral of the electric field and the polarization $ \boldsymbol{m}(\boldsymbol{r}) $ for all higher multi-moment of the electric field, yielding $ \varepsilon_{0}\boldsymbol{E}(\boldsymbol{r}) = \boldsymbol{D}(\boldsymbol{r}) - \boldsymbol{m}(\boldsymbol{r}) $ as given by \cite{bonthuis2011dielectric}
	
	\begin{equation}
	\varepsilon_{0}\boldsymbol{E}(\boldsymbol{r}) = \frac{1}{4\pi} \int \rho (\boldsymbol{r}^{'}) \frac{\boldsymbol{r} - \boldsymbol{r}^{'}}{|\boldsymbol{r} - \boldsymbol{r}^{'}|^{3}}d\boldsymbol{r}^{'}
	\end{equation}
	
	\begin{equation}
	\varepsilon_{0}\boldsymbol{E} = \boldsymbol{D} - \boldsymbol{m}
	\label{eq:6}
	\end{equation}
	where $ \rho(\boldsymbol{r}) $ is total charge density.
	Definition of total polarization $ \boldsymbol{M} $ is
	\begin{equation}
	\boldsymbol{M} = \int_{V} \boldsymbol{m}(\boldsymbol{r}) d\boldsymbol{r}
	\label{eq:7}
	\end{equation}
	The integral is over the volume of $ V $. Through fluctuation-dissipation theory, in the presence of external homogeneous electric field $ \boldsymbol{F} $, the change in the polarization is defined by \cite{bonthuis2011dielectric}
	\begin{equation}
	\Delta \boldsymbol{m}(\boldsymbol{r}) = \langle\boldsymbol{m}(\boldsymbol{r})\rangle_{\boldsymbol{F}} - \langle\boldsymbol{m}(\boldsymbol{r})\rangle_0 
	\label{eq:8}
	\end{equation}
	In the form of average on ensemble
	
	\begin{equation}
	\Delta \boldsymbol{m}(\boldsymbol{r}) = \frac{\int (\boldsymbol{m} - \langle\boldsymbol{m}\rangle_0) exp[-\beta(U - \boldsymbol{M}\cdot \boldsymbol{F})]d\mathcal{P}}{\int exp[-\beta(U - \boldsymbol{M}\cdot \boldsymbol{F})]d\mathcal{P}}
	\label{eq:9}
	\end{equation}
	where $ d\mathcal{P} = \prod_{i} d\boldsymbol{r}_i d\boldsymbol{q}_{i} $ denotes the integration on phase region over all directions and positions, indices $ 0 $ and $ \boldsymbol{F} $ denote the absence and presence of external field, respectively. This equation can be linearized for small magnitude of $ \boldsymbol{F} $ (in linear response regime) to be \cite{bonthuis2012profile}
	
	\begin{equation}
	\Delta \boldsymbol{m}(\boldsymbol{r}) \approx [ \langle\boldsymbol{m}(\boldsymbol{r})\boldsymbol{M}\rangle_{0} - \langle\boldsymbol{m}(\boldsymbol{r})\rangle_0 \langle\boldsymbol{M}\rangle_0] \cdot \boldsymbol{F}
	\label{eq:10}
	\end{equation}
	In current case study as slab geometry, ensemble averaged quantities depend only on z component of spatial region.In perpendicular direction, through vanishing of average monopole density condition as $ \nabla\cdot \boldsymbol{D}(z) = 0 $ and boundary condition of $ \Delta D_{\perp} (z) = D_{\perp} $, indicating constant displacement field. By applying  Eq.~\ref{eq:5} and Eq.~\ref{eq:6}, we reach the following equation
	\begin{equation}
	\varepsilon^{-1}_{\perp}(z) = 1 - \frac{\Delta m_{\perp}(z)}{D_{\perp}}
	\label{{eq:11}}
	\end{equation}
	we note that the field $ F_{\perp} $ is equal to $ D_{\perp}/\varepsilon_{0} $. Using equations Eq.~\ref{eq:10} and Eq.~\ref{eq:11} with the definition $ \Delta m_{\perp}(z) = \langle m_{\perp}(z)M_{\perp}\rangle - \langle m_{\perp}(z)\rangle\langle M_{\perp}\rangle $ and $ C_{\perp} = L_{x}L_{y}\int_{L_{z}}\Delta m_{\perp}(z)dz $ yields\cite{ballenegger2005dielectric}
	\begin{equation}
	\label{perpepsz}
	\varepsilon^{-1}_{\perp}(z) = 1 - \frac{\Delta m_{\perp}(z)}{\varepsilon_{0}k_{B}T + C_{\perp}/V}
	\end{equation}
	\\
	which $ L_{x} $ and $ L_{y} $ are the length of the system in x and y directions. We can calculate the polarization as an integral of the charge distribution $ \rho(\boldsymbol{r}) $ over volume in real region. Therefore, the $  m_{\perp}(z) $ could be written as \cite{schlaich2016water}
	
	\begin{equation}
	m_{\perp}(z) = -\int_{0}^{z} \rho (z^{'})dz^{'}
	\end{equation}
	$ \rho (z^{'})dz^{'} $ is the density of charges in height $ z^{'} $.\\
	
	\section{Complementary data} \label{app:complementary}
	In this part, we show the layered structure of water in confined slab geometry via density of particles and charges as a function of z in direction perpendicular to the graphene surfaces.
	\begin{figure*}[ht]
		\centering
		\includegraphics[width=1.0\linewidth]{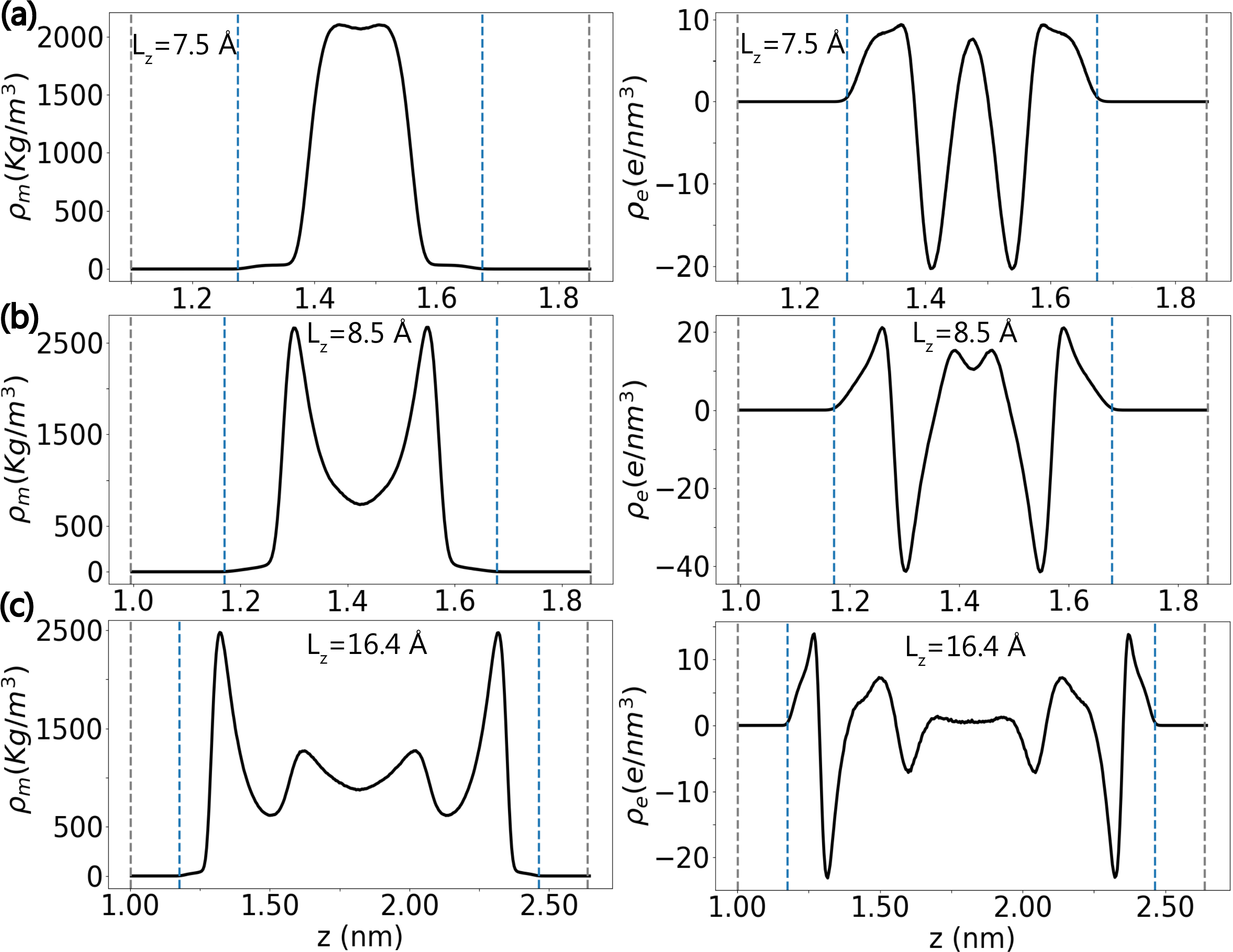}
		\caption{%
			Molecular (left) and charge (right) densities of confined water in channels with different heights. Dashed lines in gray and blue indicate the position of graphene and water, respectively. The particle density indicates the layered structure of water under confinement, it also indicates a dense layer in the first layer of water closest to graphene surfaces. Extreme reduction in the density of particles after the first apex in both sides of the diagrams shows the low-density region
			with a thickness of $\approx 2\AA$ between the first and the second layers of water. The diagrams for the density of charges reveal more tendency of hydrogen atoms to be near to graphene sheets. It
			demonstrate that the total charges in the low-density region between the first and the second layer of the water are positive, following the presence of dangling O-H bonds in these regions. The water slabs in the figure are (a) $L_{z} = 7.5\AA$, (b) $L_{z} = 8.5\AA$, and (c) $L_{z} = 16.4\AA$.
		}
		\label{fig:density-m-e}
	\end{figure*}
	
\begin{figure}[!htbp]
		\centering
		\includegraphics[width=1.0\linewidth]{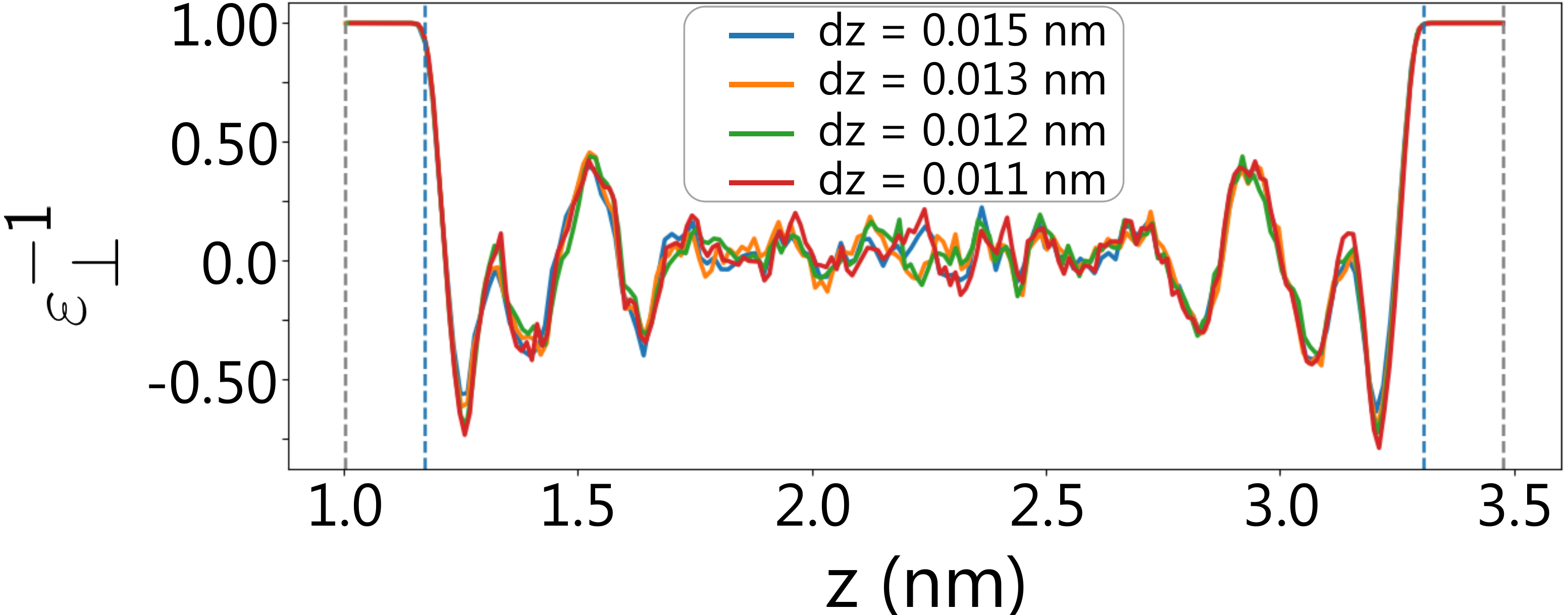}
		\caption{%
			Four different choice of dz in order to show that the changes in perpendicular component of dielectric constant is independent of bin size dz.
		}
		\label{fig:perp-248A-differentBINs}
	\end{figure}

	Here, We show the dielectric profile of water in perpendicular direction with different values of dz in order to show that the changes in perpendicular dielectric component is independent of dz value in related equations. We can see from diagram in Fig.~\ref{fig:perp-248A-differentBINs} that the trends of dielectric permittivity in $z$ direction are the same for four different choices of $dz$.
	
	Here we present more data on fitting of residence time and Eq.~\ref{eq:RTfit} for different slab thicknesses.
	
\begin{table*}[ht]
	\caption{Residence time results for \( L_{z} =12.4 \AA \) distance between graphene sheets.}
	\centering
	\begin{tabular}{ l  |   c   c   c   c   c}
		\hline
		\hline
		\( \text{Estimated Layers} \) & \( \tau_{2}(ps) \) & \( \tau_{3}(ps) \) & \( n_{2} \) & \( n_{3} \) & \( n_{p} \)  \\
		\hline
		1st layer & 19.26 & 1.42 & 441.55 & 20.6 &  \(\sim 0\)\\
		2nd layer & 8.42 & 1.05 & 360.80 & 41.82 &  \(\sim 0\)\\
		\hline
		\hline
	\end{tabular}

	\label{tab:RT-124A}
\end{table*}

\begin{table*}[ht]
	\caption{Residence time results for \( L_{z} =16.4 \AA \) distance between graphene sheets.}
	\centering
	\begin{tabular}{ l  |  l   c   c   c   c   c   c   c}
		\hline
		\hline
		\( \text{Estimated Layers} \) & \( \tau_{1}(ps) \) & \( \tau_{2}(ps) \) & \( n_{1} \) & \( n_{2} \) & \( n_{p} \)  \\
		\hline
		1st layer & 18.52 & 1.33 & 448.36 & 20.59 &  \(\sim 0\)\\
		2nd layer & 7.30 & 1.00 & 398.05 & 51.61 &  \(\sim 0\) \\
		\hline
		\hline
	\end{tabular}
	\label{tab:RT-164A}
\end{table*}

\begin{table*}[ht]
	\caption{Residence time results for \( L_{z} =42.8 \AA \) distance between graphene sheets.}
	\centering
	\begin{tabular}{ l  |  l   c   c   c   c   c   c   c }
		\hline
		\hline
		\( \text{Estimated Layers} \) & \( \tau_{1}(ps) \) & \( \tau_{2}(ps) \) & \( n_{1} \) & \( n_{2} \) & \( n_{p} \)  \\
		\hline
		1st layer & 19.84 & 1.28 & 461.9 & 18.38 & \(\sim 0\) \\
		2nd layer & 7.38 & 1.02 & 407.68 & 52.71 & \(\sim 0\)\\
		3rd layer & 4.74 & - & 421.50 & - & \(\sim 0\) \\
		Middle layer (close to bulk) & 4.27 & - & 423.13 & - & \(\sim 0\) \\
		\hline
		\hline	
	\end{tabular}
	\label{tab:RT-428A}
\end{table*}

\end{document}